\documentclass[11pt]{llncs}
\advance\hoffset by -1.6cm
\advance\voffset by -0.5cm
\usepackage{pstricks,pst-node,pst-tree,pst-plot}
\usepackage{newicktree}
\usepackage{amssymb,amsfonts,amsmath,enumerate,graphicx}
\usepackage[algo2e,boxed,vlined]{algorithm2e} 
\usepackage{url}
\renewcommand{\leq}{\leqslant}
\renewcommand{\geq}{\geqslant}
\newcommand{\TD}{\mathit{TD}}

\newcommand{\NN}{\mathbb{N}}
\newcommand{\LL}{\mathcal{L}}
\renewcommand{\SS}{\mathcal{S}}
\newcommand{\TT}{\mathcal{T}}
\renewcommand{\AA}{\mathcal{A}}
\newcommand{\length}{\mathrm{length}}

\begin{document}

\title{The transposition distance for phylogenetic trees}

\author{Francesc Rossell\'o\inst{1} \and Gabriel Valiente\inst{2}}

\institute{Department of Mathematics and Computer Science,
Research Institute of Health Science (IUNICS), University of the
Balearic Islands, E-07122 Palma de Mallorca,
\texttt{cesc.rossello@uib.es} \and Algorithms, Bioinformatics, Complexity and Formal
Methods Research Group, Department of Software,
Technical University of Catalonia, E-08034 Barcelona,
\texttt{valiente@lsi.upc.edu}}

\maketitle

\begin{abstract}
The search for similarity and dissimilarity measures on phylogenetic
trees has been motivated by the computation of consensus trees, the
search by similarity in phylogenetic databases, and the
assessment of clustering results in bioinformatics.  The transposition
distance for fully resolved phylogenetic trees is a recent addition to
the extensive collection of available metrics for comparing
phylogenetic trees.
In this paper, we generalize the transposition distance from fully
resolved to arbitrary phylogenetic trees, through a construction that
involves an embedding of the set of phylogenetic trees with a fixed
number of labeled leaves into a symmetric group and a generalization
of Reidys-Stadler's involution metric for RNA contact structures.
We also present simple linear-time algorithms for computing it.
\end{abstract}

\section{Introduction}

The need for comparing phylogenetic trees arises when alternative
phylogenies are obtained using different phylogenetic methods or
different gene sequences for a given set of species.  The comparison
of phylogenetic trees is also essential to performing phylogenetic
queries on databases of phylogenetic trees~\cite{page:2005}.  Further,
the need for comparing phylogenetic trees also arises in the
comparative analysis of clustering results obtained using different
clustering methods or even different distance matrices, and there is a
growing interest in the assessment of clustering results in
bioinformatics~\cite{handl.ea:2005}.

A number of metrics for phylogenetic tree comparison are known,
including the partition (or symmetric difference)
metric~\cite{penny.hendy:1985,robinson.foulds:1979}, the
nearest-neighbor interchange metric~\cite{waterman.smith:1978}, the
subtree transfer distance~\cite{allen.steel:2001}, the metric from the
crossover method~\cite{robinson.foulds:1981}, the quartet
metric~\cite{estabrook.ea:1985}, the metric from the nodal distance
algorithm~\cite{bluis.ea:2003}.  One of the simplest and easiest to
compute metrics proposed so far, the transposition
distance~\cite{valiente-spire2005}, is only defined for fully resolved
trees.  But phylogenetic analyses often produce phylogenies with
polytomies, that is, phylogenetic trees that are not fully resolved.
As a matter of fact, at the time of this writing, more than a 66.5\%
of the phylogenies contained in TreeBASE have polytomies.

In this paper, we generalize to arbitrary phylogenetic trees this
transposition distance, through a new definition of it.  This new
distance is directly inspired on the one hand by the matching
representation of phylogenetic
trees~\cite{diaconis.holmes:1998,stanley:1998} and on the other hand
by the \emph{involution metric} for RNA contact
structures~\cite{Reidys-Stadler96,Rossello04}.

The matching representation $M(T)$ of a phylogenetic tree $T=(V,E)$
with $n$ leaves labeled $1,\ldots,n$ describes $T$ injectively as a
partition of $\{1,\ldots,|V|-1\}$.  If $T$ is fully resolved, which is   the
particular case considered in~\cite{diaconis.holmes:1998}, then all
members of this partition are 2-elements sets, and then, since
$|V|=2n-1$, it defines an undirected 1-regular graph
$(\{1,\ldots,2n-2\},M(T))$.  Reidys and Stadler defined the
\emph{involution metric} on 1-regular graphs, by associating to each
such a graph the permutation given by the product of the transpositions
corresponding to its edges, and then using the \emph{canonical metric}
in the symmetric group $\SS_{2n-2}$ (the least number of
transpositions necessary to transform one permutation into another)
to compare these permutations.  The translation of this metric to
matching representations yields twice the matching distance defined
in~\cite{valiente-spire2005}.  Unfortunately, no meaningful
generalization to arbitrary graphs of Reidys and Stadler's metric is
known, the main drawback being the difficulty of associating
injectively a well-defined permutation to an arbitrary graph.

Now, if $T$ is not fully resolved, the members of $M(T)$ are no longer
pairs of numbers, and therefore they do not define a graph, at least
not directly.  Actually, the approach that we take in this paper can
be understood as if we represented each member
$\{i_{1},\ldots,i_{k}\}$ of $M(T)$, with $i_{1}<\cdots < i_{k}$, as a
cyclic directed graph with arcs $(i_{1},i_{2}),\ldots, (i_{k-1},i_{k}),
(i_{k},i_{1})$, and $M(T)$ as the sum of these cyclic graphs.  Now,
generalizing Reidys-Stadler's approach, we associate to every such a
cyclic directed graph the cyclic permutation $(i_{1},\ldots,i_{k})$ (if   $k=2$,
it is a transposition), and we describe $M(T)$ by means of the product
of the cyclic permutations associated to its members: since these
members are disjoint to each other, this product is
well-defined.  This defines an embedding of the set of phylogenetic
trees with $n$ leaves labeled $1,\ldots,n$ into the symmetric group
$\SS_{2n-2}$.  The transposition distance is obtained by translating
the canonical metric on $\SS_{2n-2}$ into a distance for phylogenetic
trees through this embedding.  This transposition distance measures
the least number of certain simple operations (splitting sets of
children, joining sets of children, interchanging children) that are
necessary to transform one tree into another, and it can be easily
computed in linear time.  Therefore it satisfies the requirements of
``computational simplicity'' and ``good theoretical basis'' that are
required to any distance notion on phylogenetic trees \cite{AIM03}.

\section{Matching Representation of Phylogenetic Trees}

Throughout this paper, by a \emph{phylogenetic tree} we mean a
\emph{rooted tree with injectively labeled leaves and without
outdegree 1 nodes}.  Thus, a phylogenetic tree is a directed finite
graph $T=(V,E)$ containing a distinguished node $r \in V$, called the
\emph{root}, such that for every other node $v \in V$ there exists
one, and only one, path from the root $r$ to $v$.  The \emph{children}
of a node $v$ in a tree $T=(V,E)$ are those nodes $w\in V$ such that
$(v,w) \in E$.  The \emph{outdegree} of a node is the number of its
children.  The nodes without children are the \emph{leaves} of the
tree, and the remaining nodes are called \emph{internal}: since we
assume that no node has outdegree 1, every internal node has at least
2 children.  The set of leaves of $T$ is denoted by $\LL(T)$.  The
\emph{height} of a node $v$ in a tree $T$ is the length of a longest
directed path from $v$ to a leaf.  Thus, the nodes with height 0 are
the leaves, the nodes with height 1 are the nodes all whose children
are leaves, and so on.

The leaves of a phylogenetic tree are injectively labeled in a fixed,
but arbitrary, ordered set: these labels are called \emph{taxa}.  In
practice, if the tree has $n$ leaves, we shall identify their labels
with $1,\ldots,n$, ordered in the usual increasing way.  The label
associated to a leaf $v \in V$ will be denoted by $\ell(v)$.

We shall denote by $\TT_{n}$ the set of all phylogenetic trees with
$n$ leaves labeled $1,\ldots,n$ (up to label-preserving isomorphisms of
rooted trees).

\begin{definition}
The \emph{bottom-up ordering}
(cf.~\cite{diaconis.holmes:1998,valiente:2002}) of a phylogenetic
tree $T=(V,E)\in \TT_{n}$  is
the injective mapping
$$
\ell:V \to \{1,\ldots,|V|\}
$$
defined by the following properties:
\begin{enumerate}[(a) ]
\item If $v\in \LL(T)$, then $\ell(v)$ is its label.
\item If $\mathop{height}(u) < \mathop{height}(v)$, then
$\ell(u)<\ell(v)$.
\item If $0<\mathop{height}(u) =
\mathop{height}(v)$ and
$$
\min\{\ell(x)\mid x\in\mathrm{children}(u)\} <
\min\{\ell(x)\mid x\in\mathrm{children}(v)\},
$$
then $\ell(u)<\ell(v)$.
\end{enumerate}
\end{definition}

It is straightforward to notice that this bottom-up ordering is
unique, and it can be computed in time linear in the size of the tree
by bottom-up tree traversal techniques~\cite{valiente:2002}.  First,
the leaves of $T$ are labeled by their label in $\{1,\ldots,n\}$.
Then, the height 1 nodes are labeled from $n+1$ on in the order given
by the smallest label of their children: i.e., the height 1 node with
the smallest child label is assigned label $n+1$, the height 1 node
with the next-smallest child label is assigned label $n+2$, etc.  And
this procedure is continued for consecutively increasing heights.  The
detailed pseudocode is given in Algorithm~\ref{alg:bottomupordering}.

\begin{algorithm2e}[t]
\SetFuncSty{emph}
\SetArgSty{textrm}
\dontprintsemicolon
\Begin{
\ForEach{node $v$ of $T$}{
    \eIf{$v$ is a leaf node of $T$}{
      set $\ell(v)$ to the index of $\ell(v)$ in $L$\;
    }{
      $\ell(v) := 0$\;
    }
}
$i := |L|$\;
\ForEach{level $h$ of $T$ from the leaves up to the root}{
    let $S$ be the set of nodes of $T$ at level $h$, ordered by label\;
    \ForEach{$v \in S$}{
      let $w$ be the parent of $v$ in $T$\;
      \If{$\ell(w)=0$ and $\mathop{height}(w)=h+1$}{
        $i := i+1$\;
        $\ell(w) := i$\;
      }
    }
}
return $M$\;
}
\caption{\label{alg:bottomupordering}\textbf{Bottom-up ordering}.
Given an ordered set $L$ and a  phylogenetic tree $T$ with leaves
bijectively labeled in $L$, the algorithm computes the bottom-up
ordering
of $T$.}
\end{algorithm2e}

\begin{center}
     \begin{figure}[htb]
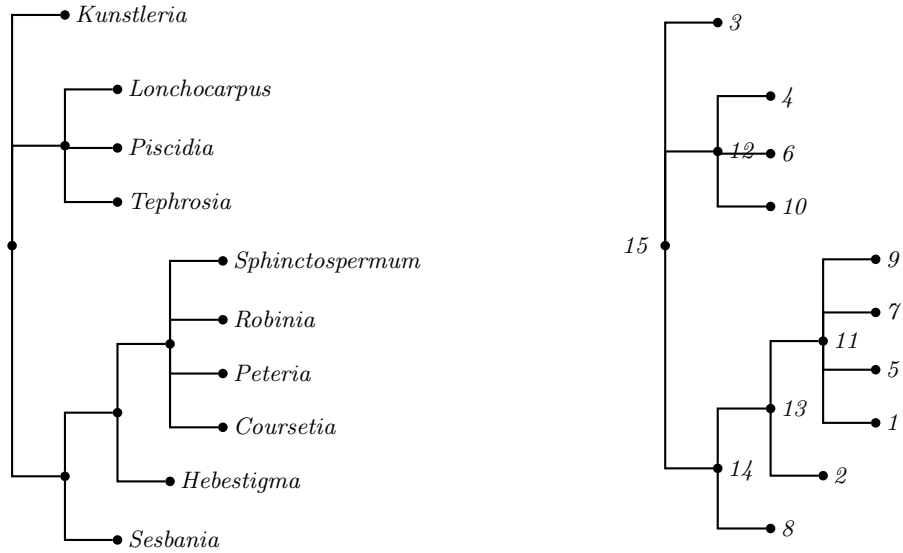

\begin{center}
{\footnotesize \begin{newicktree}
\righttree
\setunitlength{0.7cm}
\nobranchlengths
\drawtree{(Kunstleria,(Lonchocarpus,Piscidia,Tephrosia),(((Sphinctospermum,%
 Robinia,Peteria,Coursetia),Hebestigma),Sesbania));}
\qquad\qquad\qquad\qquad
\drawtree[15]{(3,(4,6,10){12},(((9,7,5,1){11},2){13},8){14});}
\end{newicktree}
}\end{center}
\caption{\label{fig:example1}%
A phylogenetic tree (left) and its bottom-up ordering (right).}
\end{figure}
\end{center}

\begin{example}\label{ex:primer}
Fig.~\ref{fig:example1} shows the Tree T166c11x6x95c08c56c38 in
TreeBASE and its bottom-up ordering after sorting its taxa
alphabetically.
\end{example}

The next definition generalizes the perfect matching representation of
binary, or fully resolved,   trees~\cite{diaconis.holmes:1998,stanley:1998}.

\begin{definition}
Let $T=(V,E)$ be a phylogenetic tree with $n$ leaves labeled
$1,\ldots,n$, and let $\ell:V\to \{1,\ldots,|V|\}$ be its bottom-up
ordering.  The \emph{matching representation} $M(T)$ of $T$ is the
partition of $\{1,\ldots,|V|-1\}$ defined as follows: $$
M(T)=\{\ell(\mathop{children}(u))\mid u\in V-\LL(T)\}.  $$
\end{definition}

\begin{example}
The matching representation of the tree in Fig.~\ref{fig:example1} is
the partition  of $\{1,\ldots,14\}$ given by
$$
\Bigl\{\{1,5,7,9\},\{4,6,10\},\{2,11\},\{8,13\},\{3,12,14\}\Bigr  \}.
$$
\end{example}

It is clear that, once the bottom-up ordering of $T$ has been
obtained, the set $M(T)$ can be produced in linear time in the size of
the tree.  Furthermore, the following two results are straightforward.

\begin{corollary}
For every $T=(V,E)\in \TT_{n}$, \mbox{$|M(T)|=|V|-n$}.
\end{corollary}

\begin{corollary}
For every $T_{1},T_{2}\in \TT_{n}$, if
$M(T_{1})=M(T_{2})$, then $T_{1}= T_{2}$.
\end{corollary}

\section{The transposition distance}

For every $m\geq 1$, let $\SS_{m}$ denote the symmetric group of
permutations of $\{1,\ldots,m\}$.  By a \emph{cycle} in $\SS_{m}$ we
understand a cyclic permutation $(i_{1},i_{2},\ldots,i_{k})\in
\SS_{m}$, with $k\geq 2$, that sends $i_{1}$ to $i_{2}$, $i_{2}$ to
$i_{3}$,\ldots, $i_{k-1}$ to $i_{k}$, and $i_{k}$ to $i_{1}$, leaving
fixed the remaining elements of $\{1,\ldots,m\}$.  Recall that the
inverse of a cycle $(i_{1},i_{2},\ldots,i_{k})$ is
$(i_{1},i_{2},\ldots,i_{k})^{-1}=(i_{k},i_{k-1},\ldots,i_{1})$: the
permutation that sends $i_{k}$ to $i_{k-1}$, $i_{k-1}$ to
$i_{k-2}$,\ldots, $i_{2}$ to $i_{1}$, and $i_{1}$ to $i_{k}$.  The
\emph{length} of a cycle $(i_{1},i_{2},\ldots,i_{k})$ is the number
$k$ of elements it moves. 

The \emph{cycle associated to a subset} $S=\{i_{1},\ldots,i_{k}\}$,
with $i_{1}<\cdots <i_{k}$ and $k\geq 2$, of $\{1,\ldots,m\}$,
is $\kappa(S):=(i_{1},i_{2},\ldots,i_{k})\in \SS_{m}$.  If
$k=1$, i.e., if $S$ is a singleton, then $\kappa(S)$ is the
identity in $\SS_{m}$, which we do not consider a cycle.

\begin{definition}
The \emph{matching permutation} $\pi(T)$ associated to a phylogenetic
tree $T=(V,E)\in \TT_{n}$ is the permutation of $\{1,\ldots,|V|-1\}$
defined by the product of the sorted cycles associated to the members of its
matching representation: 
$$
\pi(T)=\prod_{u\in V-\LL(T)} \kappa(\ell(\mathop{children}(u)))\,.
$$
\end{definition}

\begin{example}\label{ex:pit1}
The matching permutation associated to the tree in
Fig.~\ref{fig:example1} is the product of cycles
\[
(1,5,7,9)(4,6,10)(2,11)(8,13)(3,12,14)\in \SS_{14}\,,
\]
i.e., the permutation
\[
\left(
\begin{array}{cccccccccccccc}
1 & 2 & 3 & 4 & 5 & 6 & 7 &
8
& 9 & 10 & 11 & 12
& 13 & 14 \\
5  & 11 & 12 & 6 & 7 & 10 &   9
& 13
& 1 & 4 & 2 & 14
& 8 & 3
\end{array}
\right)
\]
\end{example}

If $u,v\in V-\LL(T)$ are two different internal nodes of $T$, then
$\ell(\mathop{children}(u))\cap \ell(\mathop{children}(v))=\emptyset$.
Therefore, all cycles $\kappa(\ell(\mathop{children}(u)))$ appearing
in the product defining $\pi(T)$ are disjoint to each other, and hence
they commute with each other, which implies that this product is well
defined.

Notice that no element in $\{1,\ldots,|V|-1\}$ remains fixed by
$\pi(T)$, because every $\ell(\mathop{children}(u))$, with $u$
internal, has at least two elements and every element in
$\{1,\ldots,|V|-1\}$ is the bottom-up ordering label of a child of
some internal node.  Now, if $T=(V,E)$ is a phylogenetic tree with $n$
leaves, then $|V|\leq 2n-1$, the equality holding if and only if $T$
is binary.  To be able to compare matching permutations of
phylogenetic trees with the same number of leaves $n$ but different
numbers of internal nodes, we shall understand henceforth that the   matching
permutation $\pi(T)$ belongs to $\SS_{2n-2}$, leaving fixed the
elements $|V|,\ldots,2n-2$.

The following result is a direct consequence of the facts that the
matching representation of a phylogenetic tree uniquely determines it
and every permutation has a unique decomposition as a product of
disjoint cycles of length $\geq 2$.

\begin{proposition}\label{prop:piinj}
For every $T_{1},T_{2}\in \TT_{n}$, if $\pi(T_{1})=\pi(T_{2})$, then
$T_{1}= T_{2}$.
\end{proposition}

\begin{figure}[htb]
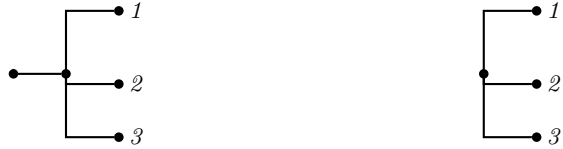

\begin{center}
{\footnotesize \begin{newicktree}
\righttree
\setunitlength{0.7cm}
\nobranchlengths
\drawtree{((1,2,3));}
\qquad\qquad\qquad\qquad\qquad\qquad
\drawtree{(1,2,3);}
\end{newicktree}
}\end{center}
\caption{\label{fig:example0}%
Two trees with the same matching permutation.}
\end{figure}

\begin{remark}
If we allow the existence of outdegree 1 nodes in our phylogenetic
trees, then the last proposition is no longer true.  Indeed, consider
the trees in Fig.~\ref{fig:example0}.  The left-hand side one has
matching representation $\{\{1,2,3\},\{4\}\}$, while the right-hand
side one has matching representation $\{\{1,2,3\}\}$.  Therefore the
matching permutation associated to both trees is $(1,2,3)$ (considered
as an element of $\SS_{4}$).
\end{remark}

Arguing as in \cite[Cor.~1]{Reidys-Stadler96}, we have the following result.

\begin{theorem}
The mapping that associates to every pair $(T_{1},T_{2})$ of
phylogenetic trees with $n$ leaves labeled in $\{1,\ldots,n\}$, the
least number $\TD'(T_{1},T_{2})$ of transpositions necessary to
represent the permutation $\pi(T_{2})^{-1}\pi(T_{1})\in\SS_{2n-2}$, is a
metric on $\TT_{n}$.
\end{theorem}

\begin{proof}
By Proposition \ref{prop:piinj}, the mapping $\pi:\TT_{n}\to
\SS_{2n-2}$ that sends every $T\in \TT_{n}$ to its matching
permutation $\pi(T)$ is an embedding.  Then, since the mapping
$$
d_{trans}:\SS_{2n-2}\times \SS_{2n-2}\to \NN
$$
defined by
$$
\begin{array}{rl}
d_{trans}(\pi_{1},\pi_{2})= & \mbox{the least number of 
transpositions necessary}\\
& \mbox{to
represent $\pi_{2}^{-1}\cdot \pi_{1}$}
\end{array}
$$
is a metric on $\SS_{2n-2}$ (see, for instance,
\cite[Thm.~2]{Reidys-Stadler96}), the mapping
$$
\begin{array}{rrcl}
\TD'& :\TT_{n}\times \TT_{n} & \to & \NN\\
& (T_{1},T_{2}) & \mapsto & d_{trans}(\pi(T_{1}),\pi(T_{2}))
\end{array}
$$
is a metric on $\TT_{n}$.\qed
\end{proof}

\begin{remark}\label{record}
Recall that the least number of transpositions required to represent a
cycle of length $k$ is $k-1$, for instance through
\[
(i_{1},\ldots,i_{k})=(i_{1},i_{2})(i_{2},i_{3})\cdots
(i_{k-1},i_{k}),
\]
and that the least number of transpositions required to represent a
product of disjoint cycles is the sum of the least numbers of
transpositions each cycle decomposes into, and hence the sum of the
cycles' lengths minus the number of cycles.
\end{remark}

The metric $\TD'$ satisfies the following property.

\begin{proposition}\label{prop:TD'parell}
For every $T_{1},T_{2}\in \TT_{n}$, $\TD'(T_{1},T_{2})$ is an even
integer smaller than $2n-2$.
\end{proposition}

\begin{proof}
If $T_{1},T_{2}\in \TT_{n}$ have $m_{1}$ and $m_{2}$ internal nodes,
respectively, then each $\pi(T_{i})$ ($i=1,2$) decomposes into $m_{i}$
disjoint cycles: say $\pi(T_{i})=C_{i,1}\cdots C_{i,m_{i}}$, with
$C_{i,j}$ of length $k_{i,j}$.  Then, by Remark~\ref{record},
$\pi(T_{i})$ has a decomposition into
$$
\sum_{j=1}^{m_{i}}(k_{i,j}-1)=\sum_{j=1}^{m_{i}} k_{i,j}-m_{i}=
n+m_{i}-1-m_{i}=n-1
$$
transpositions.  But then $\pi(T_{2})^{-1}\pi(T_{1})$ admits a
decomposition into $2(n-1)$ transpositions.  This entails that
\emph{every} decomposition of this permutation into a product of
transpositions must involve an even number of them, and therefore that
$\TD'(T_{1},T_{2})$ is an even integer.

As far as the stated upper bound for $\TD'(T_{1},T_{2})$ goes, notice
that $\pi(T_{2})^{-1}\pi(T_{1})$ moves at most $2n-2$ elements and
that if it is not the identity, then its decomposition into disjoint
cycles has at least 1 cycle. Therefore, again by
Remark~\ref{record}, a minimal decomposition of this permutation into
transpositions will involve at most $(2n-2)-1$ transpositions, and
since this number is even, this implies that $\TD'(T_{1},T_{2})\leq
2n-4$.
\end{proof}

In other words, $\TD'$ is ``artificially'' multiplied by 2.
Thus, we define a new metric on $\TT_{n}$ by dividing $\TD'$ by 2.

\begin{definition}
  The \emph{transposition distance} on $\TT_{n}$ is
  \[
  \begin{array}{rrcl}
  \TD& :\TT_{n}\times \TT_{n} & \to & \NN\\
  & (T_{1},T_{2}) & \mapsto & \frac{1}{2}\TD'(T_{1},T_{2})
  \end{array}
  \]
\end{definition}

In this way, $\TD$ takes values in $\{0,1,2,\ldots,n-2\}$.

\begin{figure}[htb]
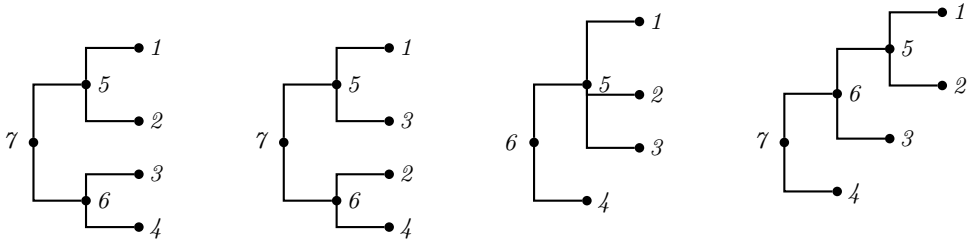

\begin{center}
{\footnotesize \begin{newicktree}
\righttree
\setunitlength{0.7cm}
\nobranchlengths
\drawtree[7]{((1,2){5},(3,4){6});}
\qquad\quad
\drawtree[7]{((1,3){5},(2,4){6});}
\qquad\quad
\drawtree[6]{((1,2,3){5},4);}
\qquad\quad
\drawtree[7]{(((1,2){5},3){6},4);}
\end{newicktree}
}\end{center}
\caption{\label{fig:example2}%
From left to right, the phylogenetic trees $T_{1}$, $T_{2}$, $T_{3}$,
and $T_{4}$ in Example~\ref{example2}.}
\end{figure}

\begin{example}\label{example2}
Let $T_{1},T_{2},T_{3},T_{4}$ be the phylogenetic trees displayed in
Fig.~\ref{fig:example2} (which we already give bottom-up ordered).
Their matching permutations are
\[
\begin{array}{l}
\pi(T_{1})=(1,2)(3,4)(5,6),\quad
\pi(T_{2})=(1,3)(2,4)(5,6),\\
\pi(T_{3})=(1,2,3)(4,5),\quad
\pi(T_{4})=(1,2)(3,5)(4,6)
\end{array}
\]
(understood as permutations in $\SS_{6}$), and then
\[
\begin{array}{rl}
\pi(T_{2})^{-1}\pi(T_{1})& =(3,1)(4,2)(6,5)(1,2)(3,4)(5,6)=(1,4)(2,3)\\
\pi(T_{3})^{-1}\pi(T_{1})& =(3,2,1)(5,4)(1,2)(3,4)(5,6)=(2,3,5,6,4)\\
\pi(T_{4})^{-1}\pi(T_{1})& =(2,1)(5,3)(6,4)(1,2)(3,4)(5,6)=(3,6)(4,5)\\
\pi(T_{3})^{-1}\pi(T_{2})& =(3,2,1)(5,4)(1,3)(2,4)(5,6)=(1,2,5,6,4)\\
\pi(T_{4})^{-1}\pi(T_{2})& =(2,1)(5,3)(6,4)(1,3)(2,4)(5,6)=(1,5,4)(3,2,6)\\
\pi(T_{4})^{-1}\pi(T_{3})& =(2,1)(5,3)(6,4)(1,2,3)(4,5)=(2,5,6,4,3)
\end{array}
\]
which yields the distances between these trees given in
Table~\ref{taula0}.
\end{example}

\begin{table}[tbp]
     \centering
     \caption{Transposition distances between pairs of trees
     $T_{1},\ldots,T_{4}$.}
     \begin{tabular}{c|c|c|c|c|}
         $\TD$ & $T_{1}$ & $T_{2}$ & $T_{3}$ & $T_{4}$  \\
         \hline
         $T_{1}$ & 0 & 1 & 2 & 1  \\
         \hline
         $T_{2}$ & 1 & 0 & 2 & 2  \\
         \hline
         $T_{3}$ & 2 & 2 & 0 & 2  \\
         \hline
         $T_{4}$ & 1 & 2 & 2 & 0  \\
         \hline
     \end{tabular}
     \label{taula0}
\end{table}

The transposition distance between two phylogenetic trees can be
easily computed in linear time.  To prove it, we move to the more
general setting of permutations and the graphs associated to them.

For every permutation $\pi\in\SS_{m}$, the \emph{directed graph}
associated to $\pi$ is the graph $G_{\pi}=(\{1,\ldots,m\},Q_{\pi})$
with
\[
Q_{\pi}=\{(i,j)\mid i\neq j\mbox{ and } \pi(i)=j\}.
\]
The directed graph $G_{\pi^{-1}}$ associated to the inverse $\pi^{-1}$
of a permutation $\pi$ is obtained by reversing all arrows in
$G_{\pi}$: thus, $Q_{\pi^{-1}}=Q_{\pi}^{-1}$ and
$G_{\pi^{-1}}=G_{\pi}^{-1}$.

Given two permutations $\pi_{1},\pi_{2}\in \SS_{m}$, by
$G_{\pi_{1}}+G_{\pi_{2}}^{-1}$ we understand the 2-colored-arcs
multigraph with set of nodes $\{1,\ldots,m\}$, set of red arcs
$Q_{\pi_{1}}$ and set of blue arcs $Q_{\pi_{2}}^{-1}$.  We shall say
that a node of $G_{\pi_{1}}+G_{\pi_{2}}^{-1}$ is \emph{unbalanced}
when it is isolated in one, and only one, of the graphs
$G_{\pi_{1}},G_{\pi_{2}}^{-1}$ (which means that it is fixed by one,
and only one, of the permutations $\pi_{1},\pi_{2}$).

\begin{proposition}\label{prop:unbalanced}
For every unbalanced node $i$ of $G_{\pi_{1}}+G_{\pi_{2}}^{-1}$:
\begin{enumerate}[(1) ]
\item If $i$ is isolated in $G_{\pi_{2}}$ and $(i_{0},i),(i,i_{1})\in
Q_{\pi_{1}}$ with $i_{0}\neq i_{1}$, then replacing the red arcs
$(i_{0},i)$ and $(i,i_{1})$ by a single red arc $(i_{0},i_{1})$
increases $d_{trans}(\pi_{1},\pi_{2})$ by 1.

\item If $i$ is isolated in $G_{\pi_{2}}$ and $(i,i_{1}),(i_{1},i)\in
Q_{\pi_{1}}$,  removing the red arcs $(i,i_{1})$ and $(i_{1},i)$
increases $d_{trans}(\pi_{1},\pi_{2})$ by 1.
\item Similar properties hold if $i$ is isolated in $G_{\pi_{1}}$ but
not in
$G_{\pi_{2}}$ and we modify the set of blue arcs.
\end{enumerate}
\end{proposition}

\begin{proof}

\noindent (1) If $(i_{0},i),(i,i_{1})\in Q_{\pi_{1}}$, with $i_{0}\neq
i_{1}$, then $i_{0}=\pi_{1}^{-1}(i)$ and $i_{1}=\pi_{1}(i)$ and hence
$(i,i_{1})\pi_{1}(i_{0})=i_{1}$, $(i,i_{1})\pi_{1}(i)=i$, and
$(i,i_{1})\pi_{1}(j)=\pi_{1}(j)$ for every $j\neq i_{0},i$.
Therefore, replacing the arcs $(i_{0},i),(i,i_{1})$ by an arc
$(i_{0},i_{1})$ is equivalent to replacing $\pi_{1}$ by
$(i,i_{1})\pi_{1}$.  So, it is enough to prove that, with the
notations and assumptions of point (1),
    $$
    d_{trans}(\pi_{1},\pi_{2})=d_{trans}((i,i_{1})\pi_{1},\pi_{2})+1.
    $$

    To prove this equality, notice that, since $i$ is fixed by $\pi_{2}$,
    $\pi_{2}^{-1}\pi_{1}$ sends $i_{0}$ to $i$ and $i$ to
    $\pi_{2}^{-1}(i_{1})$: let us denote this last index by $j_{1}$.

    If $j_{1}=i_{0}$, then $(i_{0},i)$ is a cycle of
    $\pi_{2}^{-1}\pi_{1}$ and it appears in any decomposition of this
    permutation as a product of transpositions.  But then both $i$ and
    $i_{0}$ are fixed by $\pi_{2}^{-1}((i,i_{1})\pi_{1})$, and since
    $\pi_{2}^{-1}\pi_{1}$ and $\pi_{2}^{-1}((i,i_{1})\pi_{1})$ act
    exactly in the same way on the other elements, we deduce
    that
    $$
    \pi_{2}^{-1}\pi_{1}=(i_{0},i)(\pi_{2}^{-1}((i,i_{1})\pi_{1}))
    $$
    and then
    $d_{trans}(\pi_{1},\pi_{2})=d_{trans}((i,i_{1})\pi_{1},\pi_{2})+1$
    in this
    case.

    If $j_{1}\neq i_{0}$, then the cycle of $\pi_{2}^{-1}\pi_{1}$
    moving $i_{0}$ has at least three elements:
    $$
    (i_{0},i,j_{1},j_{2},\ldots,j_{s}),\quad\mbox{ with $s\geq 1$,}
    $$
    and thus it contributes $s+1$ transpositions to a minimal decomposition
    of
    $\pi_{2}^{-1}\pi_{1}$ as a product of transpositions. Now, the cycle
    of $\pi_{2}^{-1}((i,i_{1})\pi_{1})$ that moves $i_{0}$ is
    $$
    (i_{0},j_{1},j_{2},\ldots,j_{s}),
    $$
    and it only contributes $s$ transpositions to any decomposition of
    $\pi_{2}^{-1}((i,i_{1})\pi_{1})$ as a product of transpositions.
    Therefore,
    $d_{trans}(\pi_{1},\pi_{2})=d_{trans}((i,i_{1})\pi_{1},\pi_{2})+1$
    also in this case.

    \noindent (2) If $(i,i_{1}),(i_{1},i)\in Q_{\pi_{1}}$, then
    $\pi_{1}^{-1}(i)= \pi_{1}(i)=i_{1}$, and hence
    $$
    (i,i_{1})\pi_{1}(i)=i,\quad
    (i,i_{1})\pi_{1}(i_{1})=i_{1},
    $$
    and $(i,i_{1})\pi_{1}(j)=\pi_{1}(j)$ for every $j\neq i,i_{1}$.
    Therefore, to remove the arcs $(i_{1},i),(i,i_{1})$ in this case means
    again to replace $\pi_{1}$ by $(i,i_{1})\pi_{1}$.  So, again in this
    case, it is enough to prove that, with the notations and assumptions
    of point (2),
    $$
    d_{trans}(\pi_{1},\pi_{2})=d_{trans}((i,i_{1})\pi_{1},\pi_{2})+1.
    $$

    Since $i$ is fixed by $\pi_{2}$, we have that $\pi_{2}^{-1}\pi_{1}$
    sends $i_{1}$ to $i$ and $i$ to $\pi_{2}^{-1}(i_{1})$: let us denote
    this last index by $j_{1}$.

    If $j_{1}=i_{1}$, i.e., if $i_{1}$ is also fixed by $\pi_{2}$, then
    $(i,i_{1})$ is a cycle of
    $\pi_{2}^{-1}\pi_{1}$ and it appears in any decomposition of this
    permutation as a product of transpositions.  But then both $i$ and
    $i_{1}$ are fixed by $\pi_{2}^{-1}((i,i_{1})\pi_{1})$ and
    $$
    \pi_{2}^{-1}\pi_{1}=(i_{1},i)(\pi_{2}^{-1}((i,i_{1})\pi_{1}))
    $$
    and then
    $d_{trans}(\pi_{1},\pi_{2})=d_{trans}((i,i_{1})\pi_{1},\pi_{2})+1$.

    If $j_{1}\neq i_{1}$, then the cycle of $\pi_{2}^{-1}\pi_{1}$
    moving $i_{1}$ has at least three elements:
    $$
    (i_{1},i,j_{1},\ldots,j_{s}),\quad\mbox{ with $s\geq 1$,}
    $$
    and thus it contributes $s+1$ transpositions to any decomposition of
    $\pi_{2}^{-1}\pi_{1}$ as a product of transpositions.  Now, $i$ is
    fixed by $\pi_{2}^{-1}((i,i_{1})\pi_{1})$ and the cycle of this
    permutation moving $i_{1}$ is
    $$
    (i_{1},j_{1},j_{2},\ldots,j_{s}),
    $$
    and it only contributes $s$ transpositions to any decomposition of
    $\pi_{2}^{-1}((i,i_{1})\pi_{1}))$ as a product of transpositions.
    Thus, again in this case,
    $d_{trans}(\pi_{1},\pi_{2})=d_{trans}((i,i_{1})\pi_{1},\pi_{2})+1$. \qed
\end{proof}

\begin{proposition}\label{prop:nounbalanced}
If $G_{\pi_{1}}+G_{\pi_{2}}^{-1}$ has no unbalanced node, then
\[
d_{trans}(\pi_{1},\pi_{2})= N(G_{\pi_1},G_{\pi_2}^{-1})-A(G_{\pi_1},
G_{\pi_2}^{-1}),
\]
where $N(G_{\pi_1},G_{\pi_2}^{-1})$ is the number of non-isolated
nodes of $G_{\pi_1}+G_{\pi_2}^{-1}$ and $A(G_{\pi_1},G_{\pi_2}^{-1})$
is the number of \emph{alternating cycles} in
$G_{\pi_1}+G_{\pi_2}^{-1}$, i.e., of cycles in this directed
2-colored-arcs multigraph such that two consecutive arcs have
different colors.
\end{proposition}

\begin{proof}
If $G_{\pi_1}+G_{\pi_2}^{-1}$ has no unbalanced node, then every node
either is
isolated or has exactly one incoming and one outcoming arc of each
color.  This entails that $Q_{\pi_1}\sqcup Q_{\pi_2}$ decomposes into
the union of arc-disjoint alternating cycles.

Now, every length $2k$ alternating cycle
$$
(i_{1},j_{1}),(j_{1},i_{2}),(i_{2},j_{2}),(j_{2},i_{3}),\ldots,
(i_{k},j_{k}),(j_{k},i_{1}),
$$
with $(i_{\ell},j_{\ell})\in Q_{\pi_1}$ for every $\ell=1,\ldots,k$ and
$(j_{\ell},i_{\ell+1})\in Q_{\pi_2}$ for every $\ell=1,\ldots,k-1$ and
$(j_{k},i_{1})\in Q_{\pi_2}$, corresponds to a length $k$ cycle
$$
(i_{1},i_{2},\ldots,i_{k})
$$
of $\pi_{2}^{-1}\pi_{1}$ and hence it adds $k-1$ transpositions to
any  decomposition into transpositions of this permutation.

Therefore, if we denote by $\AA(G_{\pi_1},G_{\pi_2}^{-1})$ the set of
alternating cycles in
$G_{\pi_1}+G_{\pi_2}^{-1}$, we have that
$$
\begin{array}{l}
    \displaystyle d_{trans}(\pi_{1},\pi_{2})  =
\sum_{C\in \AA(G_{\pi_1},G_{\pi_2}^{-1})}
\Bigl(\frac{\length(C)}{2}-1\Bigr)\\
\qquad \displaystyle  =
\frac{1}{2}\sum_{C\in \AA(G_{\pi_1},G_{\pi_2}^{-1})}
\length(C)-|\AA(G_{\pi_1},G_{\pi_2}^{-1})|\\
\qquad \displaystyle  =
\frac{1}{2}|Q_{\pi_1}\sqcup Q_{\pi_2}|-|\AA(G_{\pi_1},G_{\pi_2}^{-1})|.
\end{array}
$$
Finally, it is straightforward to notice that if
$G_{\pi_1}+G_{\pi_2}^{-1}$ has no unbalanced node, then
$|Q_{\pi_{1}}|=|Q_{\pi_{2}}|$ and it is equal to the number of
non-isolated nodes in this multigraph.  \qed
\end{proof}


\begin{algorithm2e}[t]
\SetFuncSty{emph}
\SetArgSty{textrm}
\dontprintsemicolon
\Begin{
Compute the bottom-up orderings of $T_1$ and $T_2$\;

\BlankLine

Compute the matching representation $M(T_{1})$ and the directed graph
$G_{1}=(\{1,\ldots,2n-2\},Q_{1})$ associated to $\pi(T_{1})$\;

\BlankLine

Compute the matching representation $M(T_{2})$ and the directed graph
$G_{2}=(\{1,\ldots,2n-2\},Q_{2})$ associated to $\pi(T_{2})^{-1}$\;

\BlankLine

$d := 0$\;

$N:=$ largest number appearing in $M(T_{1})$ or $M(T_{2})$\;

\While{$G_{1}+G_{2}$ has unbalanced nodes}{
   \ForEach{angle $\{(i_{0},i),(i,i_{1})\}$ in $Q_{1}$
with $i$ unbalanced and $i_{0}\neq i_{1}$}{
     $Q_{1}:=(Q_{1}-\{(i_{0},i),(i,i_{1})\})\cup\{(i_{0},i_{1})\}$\;
     $d:=d+1$\;
     $N:=N-1$;
   }
   \ForEach{angle $\{(i_{0},i),(i,i_{1})\}$ in $Q_{2}$
with $i$ unbalanced and $i_{0}\neq i_{1}$}{
     $Q_{2}:=(Q_{2}-\{(i_{0},i),(i,i_{1})\})\cup\{(i_{0},i_{1})\}$\;
     $d:=d+1$\;
     $N:=N-1$\;
   }
   \ForEach{$\{(i,i_{1}),(i_{1},i)\}$ in $Q_{1}$ with $i$
unbalanced}{
     $Q_{1}:=Q_{1}-\{(i,i_{1}),(i_{1},i)\}$\;
     $d:=d+1$\;
     $N:=N-1$ if $i_{1}$ is not unbalanced, $N:=N-2$ otherwise\;
   }
   \ForEach{$\{(i,i_{1}),(i_{1},i)\}$ in $Q_{2}$ with $i$
unbalanced}{
     $Q_{2}:=Q_{2}-\{(i,i_{1}),(i_{1},i)\}$\;
     $d:=d+1$\;
     $N:=N-1$ if $i_{1}$ is not unbalanced, $N:=N-2$ otherwise\;
   }
}

Compute the number $A$ of alternating cycles in the resulting directed
multigraph $G_{1}+G_{2}$, by traversing them\;

\BlankLine

$\TD(T_{1},T_{2}):=(d+N-A)/2$\;
}
\caption{\label{alg:multigraph}\textbf{Transposition distance}.
Given phylogenetic trees $T_{1},T_{2}\in \TT_{n}$, the algorithm
computes the transposition distance $\TD(T_{1},T_{2})$.}
\end{algorithm2e}

These propositions allow us to compute $\TD(T_{1},T_{2})$, for
$T_{1},T_{2}\in \TT_{n}$, in time linear on $n$ using the procedure
given in pseudocode in Algorithm~\ref{alg:multigraph}.

\begin{remark}
If $T_{1}$ and $T_{2}$ are two phylogenetic trees with different sets
of labels, then we can compute their transposition distance by first
restricting them to the sets of leaves with common labels, and then
relabeling consecutively these common labels, starting with 1.  Since
we do not allow outdegree 1 nodes, when we restrict a phylogenetic
tree to a subset of its set of taxa we contract edges to remove
outdegree 1 nodes.
\end{remark}

\begin{figure}[htb]
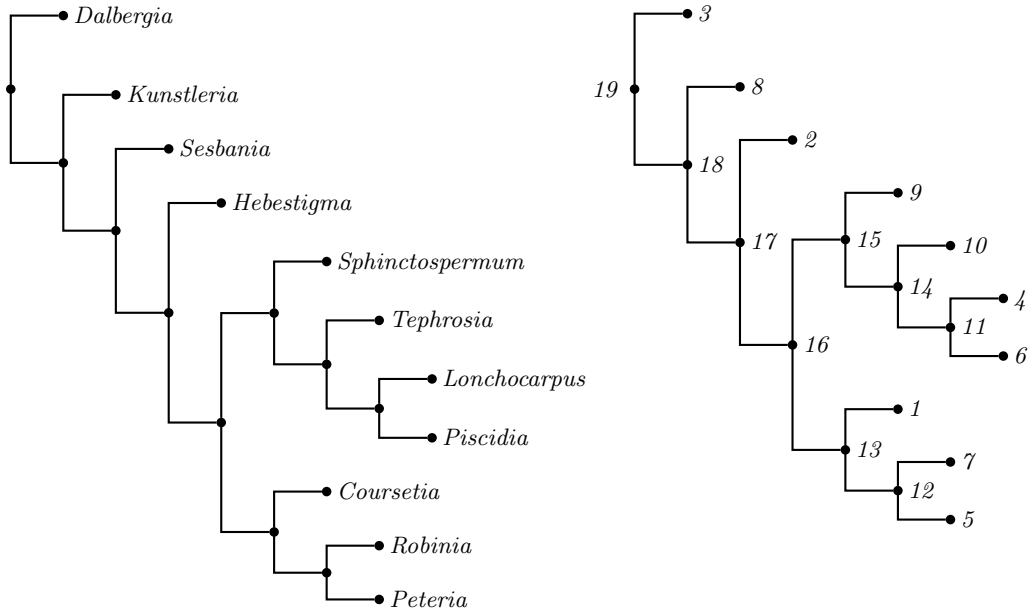

\begin{center}
{\footnotesize \begin{newicktree}
\righttree
\setunitlength{0.7cm}
\nobranchlengths
\drawtree{(Dalbergia,(Kunstleria,(Sesbania,(Hebestigma,((Sphinctospermum,(Tephrosia,(Lonchocarpus,Piscidia))),(Coursetia,(Robinia,Peteria)))))));}
\drawtree[19]{(3,(8,(2,((9,(10,(4,6){11}){14}){15},(1,(7,5){12}){13}){16  }){17}){18});}
\end{newicktree}
}\end{center}
\caption{\label{fig:nova}%
A phylogenetic tree and the botom-up ordering of its restriction to
the taxa of the tree in Fig.~\ref{fig:example1}.}
\end{figure}

\begin{example}\label{example3}
Let $T_{1}$ be the phylogenetic tree in Example \ref{ex:primer}
and let $T_{2}$ be the lower phylogenetic tree displayed in
Fig.~\ref{fig:nova}, which represents the bottom-up ordering (with its
taxa sorted alphabetically) of the tree T270c2x3x96c12c57c27 in
TreeBASE after removing the outer taxon \textit{Dalbergia} (and the
elementary root created in this way), which does not appear in
$T_{1}$.  Its matching permutation is
$$
\pi(T_{2})= (4,6)(7,5)(1,12)(10,11)(9,14)(13,15)(2,16)(8,17)(3,18).
$$
Since
$$
\pi(T_{1})=(1,5,7,9)(4,6,10)(2,11)(8,13)(3,12,14),
$$
(see Example~\ref{ex:pit1}), the multigraph
$G_{\pi(T_{1})}+G_{\pi(T_{2})}^{-1}$ has nodes $\{1,\ldots,18\}$,
red arcs
$(1,5)$, $(5,7)$, $(7,9)$, $(9,1)$, $(4,6)$, $(6,10)$, $(10,4)$, 
$(2,11)$, $(11,2)$, $(8,13)$, $(13,8)$, $(3,12)$, $(12,14)$, and $(14,3)$,
and blue arcs
$(4,6)$, $(6,4)$, $(7,5)$, $(5,7),(1,12)$, $(12,1)$, $(10,11)$,
$(11,10)$, $(9,14)$, $(14,9)$, $(13,15)$, $(15,13)$, $(2,16)$, 
$(16,2)$, $(8,17)$, $(17,8)$, $(3,18)$, and $(18,3)$.

To compute $\TD(T_{1},T_{2})$, we start with $d=0$ and $N=18$.
\begin{enumerate}
\item At the beginning, $15$, $16$, $17$ and $18$ are unbalanced.
Then, we remove the pairs of blue arcs $\{(13,15), (15,13)\}$,
$\{(2,16),(16,2)\}$, $\{(8,17),(17,8)\}$, and $\{(3,18),(18,3)\}$ and
we set $d=4$ and $N=14$.

\item In this way, the nodes $2,3,8,13$ become unbalanced.  Then, we
remove the pairs of red arcs $\{(2,11),(11,2)\}$, $\{(8,13),(13,8)\}$
and we replace the pair of red arcs $(14,3),(3,12)$  by a new
red arc $(14,12)$ and we set $d=7$ and $N=10$.

\item Now, $11$ has become unbalanced. Then, we remove the pair of
blue arcs $\{(10,11)$, $(11,10)\}$ and we set $d=8$ and $N=9$.

\item Now, $10$ has become unbalanced.  Then, we replace the pair of
red arcs $(6,10),(10,4)$ by a new red arc $(6,4)$ and we set $d=9$
and $N=8$.

\item At this moment, there does not remain any unbalanced node: the
resulting multigraph has 5 alternating cycles (a cycle
$(1,5,7,9,14,12,1)$, a cycle $(1,12,14,9,1)$, a cycle $(5,7,5)$, and
two cycles $(4,6,4)$).  Then, we have
$$
\TD(T_{1},T_{2})=\frac{1}{2}\Bigl(d+N-5\Bigr)=6\,.
$$
\end{enumerate}
\end{example}

In the Introduction we mentioned that the transposition distance
defined in this paper generalizes the transposition distance for fully
resolved trees.  This will be a direct consequence of the following
result.

\begin{proposition}
\label{thm:matching}
For every pair of binary phylogenetic trees $T_{1},T_{2}\in \TT_{n}$,
let $G=(V,E)$ be the undirected multigraph with $V=\{1,\ldots,2n-2\}$
and $E=M(T_{1}) \sqcup M(T_{2})$, and let $C$ be the set of connected
components of $G$.  Then, $\TD(T_1,T_2)=n-1-|C|$.
\end{proposition}

\begin{proof}
Let $G_{1}$ and $G_{2}$ denote the directed graphs associated to
$\pi(T_{1})$ and $\pi(T_{2})^{-1}$.  Since $T_{1}$ and $T_{2}$ are
binary, in $G_{1}+G_{2}$ for every blue or red arc $(i,j)$ there is
the inverse arc $(j,i)$ of the same color, and the graph $G$ in the
statement is the undirected graph obtained by replacing each pair of
arcs of the same color $\{(i,j),(j,i)\}$ by the undirected edge
$\{i,j\}$, which we shall understand colored with the same color as
the original pair.
    
Since $T_{1},T_{2}\in\TT_{n}$ are binary, and therefore they have
$2n-1$ nodes, no one of the $2n-2$ nodes of $G_{1}+G_{2}$ 
is unbalanced or isolated.  Then, by Proposition \ref{prop:nounbalanced},
$$
\TD(T_{1},T_{2})=\frac{1}{2}((2n-2)-A(G_1,G_2))=
n-1-\frac{1}{2}A(G_1,G_2).
$$
Moreover, $G$ is 2-regular, and therefore, every connected component
in $G$ is an alternating cycle, which contains exactly two 
alternating cycles of $G_{1}+G_{2}$. Therefore
$A(G_1,G_2)=2|C|$. Combining this equality with the expression for 
$\TD(T_{1},T_{2})$ given by Proposition \ref{prop:nounbalanced}, we obtain the expression in the 
statement. \qed
\end{proof}

In~\cite{valiente-spire2005}, the \emph{transposition distance}
between two \emph{binary phylogenetic trees} $T_{1}$ and $T_{2}$ was
defined as the least number of transpositions necessary to transform
$M(T_{1})$ into $M(T_{2})$: in this context, a \emph{transposition}
means a replacement of a pair of 2-elements sets $\{i,j\},\{k,l\}$ by
a new pair $\{i,k\},\{j,l\}$.  Theorem 1 in \textsl{loc.\ cit.} and
the last proposition entail that, for binary phylogenetic trees, our
transposition distance and the transposition distance defined
in~\cite{valiente-spire2005} are the same.

\section{Results}

We have implemented in Perl the algorithms for the transposition
distance between phylogenetic trees, using the BioPerl collection of
Perl modules for computational biology~\cite{stajich.ea:2002}.  The
software is available in source code form for research use to
educational institutions, non-profit research institutes, government
research laboratories, and individuals, for non-exclusive use, without
the right of the licensee to further redistribute the source code.
The software is also provided for free public use on a web server, at
the address \texttt{http://www.lsi.upc.edu/\~{}valiente/}

Using this implementation, we have performed a systematic study of the
TreeBASE~\cite{morell:1996} phylogenetic database, the main
repository of published phylogenetic analyses, which currently
contains 2,592 phylogenies with 36,593 taxa among them.  Previous
studies have revealed that TreeBASE constitutes a scale-free
network~\cite{piel.ea:2003}.

\begin{figure}
\begin{center}
\includegraphics{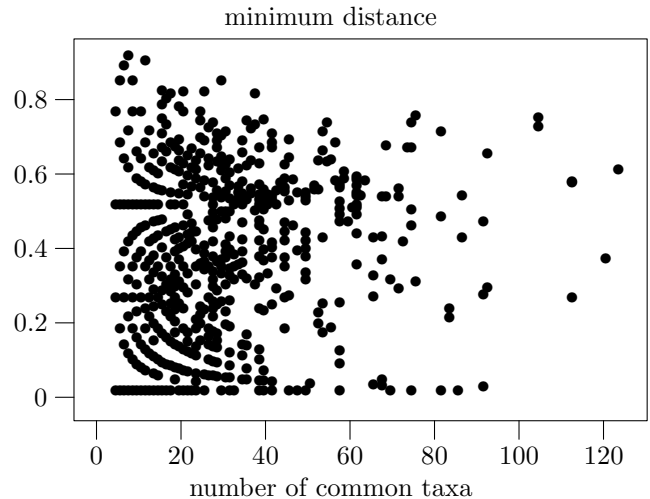}
\end{center}
\caption{\label{fig:treebase}%
Similarity of phylogenetic trees in TreeBASE based on the transposition distance. Each bullet represents the distance between a phylogenetic tree and the most similar phylogenetic tree in TreeBASE (other than itself) with at least three common taxa.}
\end{figure}

In order to assess the usefulness of the new distance measure in
practice, we have computed the transposition distance for each of the
$2,592 \cdot 2,591 / 2 = 3,357,936$ pairs of phylogenetic trees in
TreeBASE. Then, for each phylogenetic tree, we have recovered the most
similar phylogenetic tree in TreeBASE (other than itself) with at least three taxa in
common.  The results, summarized in Fig.~\ref{fig:treebase}, show that
the transposition distance allows for a good recall of similar
phylogenetic trees.

\section*{Acknowledgements}
This work has been partially supported by the Spanish DGES project
BFM2003-00771 ALBIOM, the Spanish CICYT project TIN 2004-07925-C03-01
GRAMMARS, and the UE project INTAS IT 04-77-7178.

\end{document}